\documentclass[lettersize,twoside,journal]{IEEEtran}

\usepackage[utf8]{inputenc}    
\usepackage{amsmath,amssymb,amsfonts,amsthm} 
\usepackage{graphicx}          
\usepackage[table]{xcolor}     
\usepackage{booktabs}          
\definecolor{rowblue}{RGB}{214, 228, 248}
\usepackage{braket}            
\usepackage{soul}              
\usepackage{enumerate}         
\usepackage{multirow}          
\usepackage{array}             
\usepackage{subfigure}         
\usepackage{afterpage}         
\usepackage{fancyhdr}          
\usepackage{fancyvrb}          
\usepackage{siunitx}
\usepackage{acro}
\usepackage{cite}
\usepackage{physics}

\usepackage{balance}
\usepackage{hyperref}
\usepackage{color}
\usepackage{dsfont}
\usepackage{bbm}

\DeclareMathAlphabet{\mathsf}{OML}{cmbr}{m}{it}

\newtheorem{property}{Property}

\newcommand{\HS}{\mathcal{H}}
\newcommand{\dmS}{\mathcal{D\left( \HS\right)}}
\newcommand{\boS}{\mathcal{B\left( \HS\right)}}

\newcommand{\QCg}[1]{\mathcal{N}\left( #1\right)}
\newcommand{\QCs}[2]{\mathcal{N}_{\mathrm{#1}}\left( #2\right)}

\newcommand{\QCN}{\mathcal{N}}

\newcommand{\mc}[1]{\mathcal{#1}}

\DeclareAcronym{BB84}{
	short = BB84,
	long = Bennett-Brassard 1984,
	first-style = short
}

\DeclareAcronym{BSC}{
	short = BSC,
	long = binary symmetric channel
}

\DeclareAcronym{CDC}{
	short = CDC,
	long = cooperative dense coding
}

\DeclareAcronym{FRQI}{
	short = FRQI,
	long = flexible representation of quantum images
}

\DeclareAcronym{GQIR}{
	short = GQIR,
	long =	generalized quantum image representation
}

\DeclareAcronym{NEQR}{
	short = NEQR,
	long =	novel enhanced quantum representation
}

\DeclareAcronym{PTM}{
	short = PTM,
	long = Pauli transfer matrix
}

\DeclareAcronym{POVM}{
	short = POVM,
	long = positive operator-valued measure
}

\DeclareAcronym{QBER}{
	short =QBER,
	long = quantum bit error rate
}

\DeclareAcronym{QRAC}{
	short = QRAC,
	long = quantum random access code
}

\DeclareAcronym{QKD}{
	short = QKD,
	long = quantum key distribution
}

\DeclareAcronym{QPIE}{
	short = QPIE,
	long = quantum probability image encoding
}

\DeclareAcronym{QSQI}{
	short = QSQI,
	long = quantum state with color information
}

\DeclareAcronym{SBM}{
	short = SBM,
	long = sequential basis matching
}

\DeclareAcronym{SS}{
	short = SS,
	long = six-state,
	first-style = short
}

\DeclareAcronym{SQC}{
	short = SQC,
	long = symmetric quantum cloning
}

\DeclareAcronym{SSIM}{
	short = SSIM,
	long = structural similarity index measure
}

\DeclareAcronym{SVD}{
	short = SVD,
	long = singular value decomposition
}

\DeclareAcronym{UQCM}{
	short = UQCM,
	long = universal quantum cloning machine
}

\begin{document}

\title{Quantum Key Distribution Without Shared Reference Frame Under Unital Noise}


\author{Junaid ur Rehman, Shehbaz Tariq, and Symeon Chatzinotas, \IEEEmembership{Fellow, IEEE}\\
	
	\thanks{
%
		J.~ur~Rehman is with the Department of Electrical Engineering, the Interdisciplinary Research Center for Advanced Quantum Computing,
		and the Interdisciplinary Research Center for Communication and System Sensing, King Fahd University of Petroleum and Minerals (KFUPM), Dhahran 31261, Saudi Arabia (e-mail:junaid.urrehman@kfupm.edu.sa). S.~Tariq,
		and S.~Chatzinotas 
		are with Interdisciplinary Centre for Security, Reliability and Trust (SnT), University of Luxembourg, 1855 Luxembourg City, Luxembourg (e-mail: shehbaz.tariq@uni.lu; symeon.chatzinotas@uni.lu). 
        \\
        J.~ur~Rehman would like to acknowledge the support from the KFUPM through the Ibn Battuta Global Scholarship grant
        with number ISP2651. The work of Shehbaz~Tariq and Symeon~Chatzinotas was supported by the project LUQCIA Funded by the European Union – Next Generation EU, with the collaboration of the Department of Media, Connectivity and Digital Policy of the Luxembourgish Government in the framework of the RRF program.
        \textit{(Corresponding author: Junaid ur Rehman.)}
		%
	}

}
\markboth{
	Submitted to arXiv on \today.}{ 
	ur Rehman:
	Quantum Key Distribution Without Shared Reference Frame Under Unital Noise
}

\maketitle

\begin{abstract}
We consider a general and practical scenario of \ac{QKD} over an unknown, stationary, unital qubit channel. Furthermore, due to practical limitations, e.g., relative movement and rotation of communicating parties, a global shared reference frame cannot be established. This scenario can routinely appear in satellite \ac{QKD}. We propose two methods to overcome the physical qubit noise and the lack of shared reference frame. The first proposed approach involves constructing the \ac{PTM} description of the channel, which we achieve without requiring a shared reference frame, by absorbing the lack of shared reference frame in the channel definition. This is followed by the identification of singular vectors of \ac{PTM} as the Bloch vectors for optimal signal states. In the optimized local bases, the resulting correlations are equivalent, up to outcome relabeling, to those of a Pauli channel, allowing us to show the optimality of the BB84 and six-state \ac{QKD} protocols under these conditions. The second approach, called the \ac{SBM} involves sequentially identifying the channel-optimized local bases that enable \ac{QKD}. We show that both of these approaches result in the same effective key exchange rate for \ac{QKD}.
\end{abstract}
\acresetall


\section{Introduction}

\Ac{QKD} enables two distant parties to establish secret keys with security based on quantum-mechanical principles rather than computational hardness assumptions \cite{Scarani:secu:09, Gisin:Quan:02, Pirandola:Adva:20, Portmann:Secu:22}. This distinction matters for long-term confidentiality, since information-theoretically secure keys are not compromised by future increases in computational power. Practical discrete-variable protocols such as \ac{BB84} \cite{Bennett:Quan:14, Bennett:Quan:84} and six-state \ac{QKD} \cite{Bruss:Opti:98} encode qubits in mutually unbiased preparation and measurement bases. Their achievable key rates depend directly on the \acp{QBER} observed in those bases. Therefore, both the security analysis and practical performance of \ac{QKD} depend on how accurately the physical system realizes the assumed bases and their relative alignment.

Satellite \ac{QKD} is a promising pathway toward global-scale quantum-secure communication because free-space optical links can mitigate the distance limitations imposed by fiber attenuation. Notably, the Micius quantum experiments in space \cite{Lu:Mici:22} and the experimental quasi-single-photon transmission from satellite to Earth \cite{Yin:Expe:13} demonstrate the feasibility of long-distance satellite-ground quantum optical links. However, satellite \ac{QKD} also introduces practical challenges that are less prominent in fixed terrestrial links, including time-varying link geometry, acquisition-pointing-tracking constraints, atmospheric loss, detector noise, platform motion, and attitude drift \cite{Lu:Mici:22, Yin:Expe:13, Diamanti:Prac:16}. For polarization-encoded satellite \ac{QKD}, an important additional challenge is maintaining a common polarization reference frame between the transmitter and receiver \cite{klicnik_real-time_2026}.

Standard \ac{BB84} and six-state \ac{QKD} implicitly assume that Alice's preparation bases and Bob's measurement bases are aligned. 
For \ac{BB84}, this means that the two bases used for key generation and parameter estimation correspond to the same physical directions at Alice and Bob. For the six-state protocol, the assumption is stronger, since the full set of three mutually unbiased qubit bases must be consistently aligned between the two parties. Without a shared reference frame, Alice's and Bob's local qubit descriptions may be related by an unknown rotation, so a state prepared in one local basis may not be measured in the intended corresponding basis at the receiver \cite{Bartlett:Clas:03, vanEnk:Quan:06}. This problem  appears naturally in polarization-encoded satellite links and in other platforms where the relative orientation between communicating parties is not fixed \cite{Laing:Refe:10}.

Several approaches have been proposed to mitigate reference-frame mismatch. In satellite links, polarization-basis tracking can be supported by orbit prediction and optical compensation, for example using an autorotatable half-wave plate \cite{Lu:Mici:22,Yin:Expe:13}. Reference-frame-independent \ac{QKD} also reduces the need for active alignment by using measurement statistics that are insensitive to certain frame rotations \cite{Laing:Refe:10}. These approaches address important aspects of basis mismatch. However, when the physical qubit channel is also unknown, the observed statistics reflect both reference-frame misalignment and physical qubit noise. In this setting, it is not enough to consider only how to restore a common reference frame; one must also determine how Alice and Bob should choose their local \ac{QKD} bases when neither the reference frame nor the channel's preferred noise directions are known.

In this work, we consider an unknown, stationary, unital qubit channel. Unital channels preserve the maximally mixed state and include important qubit noise models such as random unitary channels and Pauli channels \cite{Kraus:Stat:83, Nielsen:Quan:10, Wilde:Quan:17}. This assumption is natural for polarization-based optical communication when the dominant impairments are random polarization transformations, depolarization, or anisotropic Pauli-type polarization noise. However, a Pauli-channel description normally assumes that the Pauli error directions are defined with respect to a known reference frame \cite{Rehman:Enta:21}. When no shared reference frame is available, the channel may still have preferred noise directions, but Alice and Bob do not know how those directions are oriented relative to their local Pauli frames.

Related works address parts of this setting from different directions. Communication without a shared reference frame has been studied in \cite{Bartlett:Clas:03, vanEnk:Quan:06}, and reference-frame-independent \ac{QKD} was proposed in \cite{Laing:Refe:10}. Mismatched-basis statistics and tomography have been used to improve \ac{QKD} analysis or relax assumptions on the source and channel \cite{Yin:Mism:14, Watanabe:Tomo:08}. \ac{QKD} under asymmetric noise has also been analyzed, showing that basis-dependent \acp{QBER} can strongly affect the achievable key rate \cite{Murta:Key:20}. The problem addressed here is the joint one: how to identify channel-adapted local bases for \ac{BB84} and six-state \ac{QKD} when both the reference-frame mismatch and the stationary unital qubit channel are unknown.

Our main idea is to treat the unknown reference-frame mismatch as part of the channel observed by Alice and Bob. Rather than separately estimating Alice's local frame, Bob's local frame, and the physical noise process, the parties characterize the effective transformation from Alice's local preparations to Bob's local measurements. This effective-channel viewpoint allows the reference-frame mismatch and the unital qubit noise to be handled jointly, using only experimentally accessible preparation-and-measurement statistics. The formal channel model and its \ac{PTM} representation are introduced in the following sections.

Using this effective-channel description, we develop optimized local signaling bases for \ac{QKD}. One method reconstructs the effective \ac{PTM} and uses \ac{SVD} to identify the local preparation and measurement directions that best match the channel. A second method, called \ac{SBM}, finds the same bases operationally through a sequential search over mutually unbiased directions. In the optimized bases, the observed correlations are equivalent, up to outcome relabeling, to those of a Pauli channel. The resulting asymmetric \acp{QBER} are then used to evaluate the asymptotic key rates of \ac{BB84} and six-state \ac{QKD}.

We also propose a second method, called \ac{SBM}, which provides an operational alternative to explicit \ac{PTM} reconstruction and decomposition. Instead of first estimating the effective \ac{PTM}, \ac{SBM} adaptively searches for the local preparation and measurement bases that maximize the probability that Bob obtains the measurement outcome associated with Alice's prepared state. Equivalently, the search identifies bases that minimize the corresponding basis-dependent \ac{QBER}. The procedure identifies one optimized basis at a time while preserving the mutually unbiased structure required for \ac{BB84} and six-state \ac{QKD}. Under exact channel estimation and successful convergence of the search, \ac{SBM} yields the same optimized \acp{QBER} and asymptotic secret key rates as the \ac{PTM}/\ac{SVD}-based method.

The key contributions of this paper are:
\begin{itemize}
    \item We formulate \ac{QKD} without a shared reference frame over an unknown, stationary, unital qubit channel, motivated by satellite \ac{QKD} and other moving-platform quantum communication scenarios. We show that the reference-frame mismatch and physical qubit noise can be absorbed into a single effective Alice-to-Bob channel.

    \item We develop two basis-optimization methods, namely the \ac{PTM}/\ac{SVD} method and \ac{SBM}, for identifying local mutually unbiased preparation and measurement bases adapted to this effective channel.

    \item We show that, in the optimized bases, the nontrivial Bloch block of the effective channel is diagonal, so the resulting correlations are equivalent, up to outcome relabeling, to those of a Pauli channel.

    \item We evaluate \ac{BB84} and six-state \ac{QKD} under the resulting asymmetric \acp{QBER} and show through numerical examples that optimized signaling can recover positive asymptotic key rates in regimes where standard local-basis signaling may fail.
\end{itemize}

The remainder of this paper is organized as follows. 
Section~\ref{sec:preliminaries} presents the required preliminaries and system model. 
Section~\ref{sec:opt_QKD} develops the proposed optimized-signaling methods based on \ac{PTM}/\ac{SVD} and \ac{SBM}. 
Section~\ref{sec:NE} provides numerical examples for random unitary channels and compares optimized signaling with standard \ac{BB84} and six-state \ac{QKD}. 
Section~\ref{sec:conclusion} concludes the paper and outlines future directions.
\begin{table}[t]
\caption{List of Notations and Symbols}
\label{tab:symbols}
\centering
\footnotesize
\renewcommand{\arraystretch}{1.18}
\setlength{\tabcolsep}{3pt}
\begin{tabular}{@{}>{\centering\arraybackslash}p{0.30\linewidth}
                >{\raggedright\arraybackslash}p{0.64\linewidth}@{}}
\toprule
\textbf{Symbol} & \multicolumn{1}{c}{\textbf{Definition}} \\
\midrule

\rowcolor{rowblue}
\multicolumn{2}{c}{\textbf{Hilbert Space and States}} \\
$\HS$ & Hilbert space of a qubit \\
$\mathcal{D}(\HS)$ & Convex set of density operators on $\HS$ \\
$\mathcal{B}(\HS)$ & Set of bounded operators on $\HS$ \\
$\rho$ & Density operator \\
$\ket{\psi}$ & Pure state vector \\
$\vec{r}$ & Pauli state vector of a qubit state; its nonidentity components form the Bloch vector \\
\midrule

\rowcolor{rowblue}
\multicolumn{2}{c}{\textbf{Pauli Operators and Pauli Transfer Matrix}} \\
$I,\, X,\, Y,\, Z$ & Identity operator and Pauli operators \\
$P_j$ & Pauli basis operator, with $P_0=I$, $P_1=X$, $P_2=Y$, and $P_3=Z$ \\
$R^{\mc{N}}$ & \ac{PTM} representation of the channel $\mc{N}$ \\
$R^{\mc{N}}_{i,j}$ & $(i,j)$-th entry of the \ac{PTM} of channel $\mc{N}$ \\
$R^{A \to B}$ & Effective \ac{PTM} from Alice's local frame to Bob's local frame \\
\midrule

\rowcolor{rowblue}
\multicolumn{2}{c}{\textbf{Quantum Channel and Reference Frames}} \\
$\mc{N}$ & Unital qubit channel \\
$K_i$ & Kraus operators of $\mc{N}$ \\
$U_A,\, U_B$ & Unknown local-frame unitaries associated with Alice and Bob \\
$p_I,\, p_X,\, p_Y,\, p_Z$ & Pauli-channel error probabilities \\
\midrule

\rowcolor{rowblue}
\multicolumn{2}{c}{\textbf{SVD and Basis Optimization}} \\
$O_A,\, O_B$ & Orthogonal matrices whose columns define Alice's preparation directions and Bob's measurement directions, respectively \\
$\Sigma$ & Diagonal matrix of singular values obtained from the \ac{SVD} step applied to $R^{A \to B}$ \\
$\sigma_1 \geq \sigma_2 \geq \sigma_3$ & Ordered nontrivial singular values used to define the optimized basis-dependent \acp{QBER} \\
$\theta,\, \phi$ & Polar and azimuthal angles parameterizing a pure qubit state \\
\midrule

\rowcolor{rowblue}
\multicolumn{2}{c}{\textbf{QKD Performance}} \\
$Q_z,\, Q_x,\, Q_y$ & \acp{QBER} in the $Z$, $X$, and $Y$ bases \\
$h(x)$ & Binary Shannon entropy \\
$H(\{p_i\})$ & Shannon entropy, $-\sum_i p_i \log_2 p_i$ \\
$R_{\mathrm{BB84}}$ & Asymptotic \ac{BB84} secret key rate in bits per sifted detected signal \\
$R_{\mathrm{six\text{-}state}}$ & Asymptotic six-state \ac{QKD} secret key rate in bits per sifted detected signal \\
$\lambda_{i,j}$ & Eigenvalue parameters in the six-state key-rate expression \\
\bottomrule
\end{tabular}
\end{table}

\section{Preliminaries \& System Model}\label{sec:preliminaries}

\subsection{Preliminaries}
A quantum state $\rho$ is a unit-trace positive semidefinite operator, i.e., density operator, on the Hilbert space $\HS$. We denote by $\dmS$, the convex set of density operators. The extremal points of this set are the pure states $\rho = \ketbra{\psi}$ and can be equivalently represented by the state vector $\ket{\psi}\in\HS$. A quantum state $\rho$ can be decomposed in Pauli basis
\begin{align}
	\rho = \frac{1}{2}\left( I + r_x X + r_Y Y + r_z Z\right),
\end{align}
where $I$ is the identity matrix and 
\begin{align}
	X = \begin{bmatrix}
		0 & 1 \\
		1 & 0
	\end{bmatrix}, \quad	
	Y = \begin{bmatrix}
		0 & -i \\
		i & 0
	\end{bmatrix}, \quad \text {and }	
	Z = \begin{bmatrix}
		1 & 0 \\
		0 & -1
	\end{bmatrix}	
\label{eq:Pauli_operator}
\end{align}
are the well-known Pauli matrices. The vector $\vec{r} = \left[1, r_X, r_Y, r_Z\right]$ is called the Pauli state-vector \cite[Suppelemntal Material]{Chow:Univ:12}. 

 A quantum channel $\QCg{\cdot}$ is a trace-preserving completely positive map
\begin{align}
	\QCN \colon \boS \mapsto \boS,
\end{align}
where $\boS$ is the set of bounded operators on $\HS$. Clearly $\dmS \subset \boS$. A convenient representation of quantum channels is the well-known Kraus operator-sum representation \cite{Kraus:Stat:83, Nielsen:Quan:10}
\begin{align}
	\QCg{\rho} = \sum_{i}K_i \rho K_i^{\dagger},
\end{align}
where the Kraus operators $K_i$ satisfy $\sum_i K_i^{\dagger}K_i=I$. In this work, we are particularly interested in \emph{unital} quantum channels that are defined by their property $\QCg{I} = I$, by satisfying $\sum_i K_i K_i^{\dagger} = I$. Physically, these maps do not ``unmix'' a mixed density operator \cite{Nielsen:Gate:21}. Special examples of unital maps include random unitary channels, $K_i = \sqrt{p_i}U_i$ with unitaries $U_i$ and probability vector $\left[ p_i\right]_i$. A Pauli channel is a special case of the random unitary channel where the channel unitaries are the Pauli operators \eqref{eq:Pauli_operator}:
\begin{align}
	\QCs{P}{\rho} = p_I \rho + p_X X \rho X^{\dagger} + p_Y Y \rho Y^{\dagger} + p_Z Z \rho Z^{\dagger}.
	\label{eq:Pauli_channel}
\end{align}

The overlap of a quantum state $\rho$ with another state $\sigma$ is given by $\tr(\rho \sigma)$. In case of at least one of $\rho$ or $\sigma$ being pure, this quantity is same as the state fidelity and can be interpreted as the component of one of the states in the direction of the other. In the communication scenario, this can also be interpreted as the probability that the state is $\rho$ would pass a test for being the same as $\sigma$ or vice versa \cite{Wilde:Quan:17}. The following elementary property of unital channels will be helpful in our later discussion.  

\begin{property}
	\label{property:unital_equal_overlap}
	For two orthogonal qubit states $\ket{\psi_0}$, $\ket{\psi_1}$, and the unital channel $\QCN$
	\begin{align}
		\Tr{\ketbra{\psi_0}\QCg{\ketbra{\psi_1}}} = \Tr{\ketbra{\psi_1}\QCg{\ketbra{\psi_0}}}.
	\end{align}
\end{property}
\begin{proof}
	We have 
	\begin{align}
		&\Tr{\ketbra{\psi_0}\QCg{\ketbra{\psi_1}}}\\
		&= \Tr{\left(I - \ketbra{\psi_1}\right) \QCg{I - \ketbra{\psi_0}}}\\
		&= \Tr{\left(I - \ketbra{\psi_1}\right) \left(\QCg{I} - \QCg{\ketbra{\psi_0}}\right)}\\
		&= \Tr{\ketbra{\psi_1}\QCg{\ketbra{\psi_0}}}.
	\end{align}
	The first equality is due to the completeness of the orthonormal basis, i.e., $\ketbra{\psi_0} + \ketbra{\psi_1} = I$. The second equality is due to the linearity of $\QCN$. The last equality is due to the linearity of trace operator, the unital property of the channel $\QCg{I} = I$,  and the fact that the trace of a density operator is unity. 
\end{proof}
This means that the component of $\ket{\psi_0}$ introduced by the unital channel when acting on $\ket{\psi_1}$ is the same the other way round as well. Communicating classical/digital data over quantum channels often employs such sets of orthogonal states for communication. Then, we can define and interpret $p^e_{x, y}= \Tr{\ketbra{\psi_x}\QCg{\ketbra{\psi_y}}}, $ $x \neq y$ as the probability of error when states $\left\{\ket{\psi_0}, \ket{\psi_1}\right\}$ are employed for communication and projected in the same basis on the decoder's side. The Property~\ref{property:unital_equal_overlap} establishes that the errors are symmetric, i.e., $p^e_{0, 1} = p^e_{1, 0} = \epsilon$, and this setting of basis states as input and projection on the same basis at the output simulates a \emph{binary symmetric channel}.

\subsection{Pauli Transfer matrix}
An equivalent representation of quantum channels is via \ac{PTM} \cite{Chow:Univ:12}. The entries of \ac{PTM} of a quantum channel $\QCN$ are defined as $R^{\QCN}_{i, j} = \frac{1}{2}\Tr{P_i \QCg{P_j}}$, where $P_0 = I$ and $P_1, P_2, P_3$ are the Pauli matrices $X$, $Y$, and $Z$, respectively. The \ac{PTM} representation makes several of the quantum channel properties explicit and simple \cite{Merkel:Self:13, Greenbaum:Intr:15}. For example, the map $\QCN$ is trace preserving if and only if the first row of $R^{\QCN}$ is the vector $\left[1, 0, 0, 0\right]$ \cite{Merkel:Self:13}. Similarly, the map is unital if the first column has the same structure. 

\begin{property}[\cite{Merkel:Self:13}]\label{property:composition}
	The composition of multiple channels become matrix multiplication of their respective \acp{PTM}, i.e., if two quantum channels operate one after the other $\mc{N}\circ \mc{M} \left( \cdot \right)$, then the \ac{PTM} of the composite channel is the matrix multiplication of two \acp{PTM}, i.e., $R^{\mc{N}\circ \mc{M}} = R^{\mc{N}} R^{\mc{M}}$.
\end{property}

A powerful consequence of the \ac{PTM} representation of quantum channels is the ability to construct the description of the channel with fewer experimental configurations if some knowledge/probable assumption about the channel is known. This is in contrast to general quantum process tomography where $d^4$ experimental configurations are needed to obtain the complete classical description of a $d$-dimensional channel \cite{Roncallo:Paul:24}. For example, the direct reconstruction of \ac{PTM} of a $d$-dimensional unital channel can be achieved with $\left(d^2 - 1\right)^2$ experimental configurations \cite{Roncallo:Paul:24}. Indeed, with some effort we can see that for $i, j \neq 0$ we can simplify $R^{\QCN}_{i, j} = \Tr{P_i \QCg{\ketbra{\lambda_j}}}$ for a unital $\QCN$, where $\ket{\lambda_j}$ is the eigenstate of $P_j$ with eigenvalue +1. Thus, each nontrivial entry of \ac{PTM} can be estimated with a single experimental configuration. 

It can be verified that for the case of Pauli channel, the \ac{PTM} is diagonal with entries
\begin{align}
	R_{0, 0} &= 1\\ 
	R_{1, 1} &= p_I + p_X - p_Y - p_Z\\
	R_{2, 2} &= p_I - p_X + p_Y - p_Z\\
	R_{3, 3} &= p_I - p_X - p_Y + p_Z.
\end{align}
We can observe that the last these entries are related to the \acp{QBER} when using the $X$, $Y$, and $Z$ basis states for classical communication under Pauli channel \cite{urRehman:Hole:18}. Indeed, the error rate in the $X$ basis 
\begin{align}
Q_x = p_Y + p_Z = \frac{1 - R_{1, 1}}{2},
\end{align}
where we have used the fact that $p_I + p_X + p_Y + p_Z = 1$. 
Similarly, we can write
\begin{align}
	Q_y &= p_X + p_Z = \frac{1 - R_{2, 2}}{2}\\
	Q_z &= p_X + p_Y = \frac{1 - R_{3, 3}}{2}.
\end{align}

Finally, the \ac{PTM} enables the direct mapping between the Pauli state-vectors of channel input and output. That is, the Pauli state-vector $\vec{s}$ of channel output is related to the Pauli state-vector $\vec{r}$ by the \ac{PTM} $R$ of the channel: $\vec{s} = R\,\vec{r}$ \cite[Supplementary Material]{Chow:Univ:12}.


\subsection{System Model}
\begin{figure}[t]
	\centering
	\includegraphics[width=0.275\textwidth]{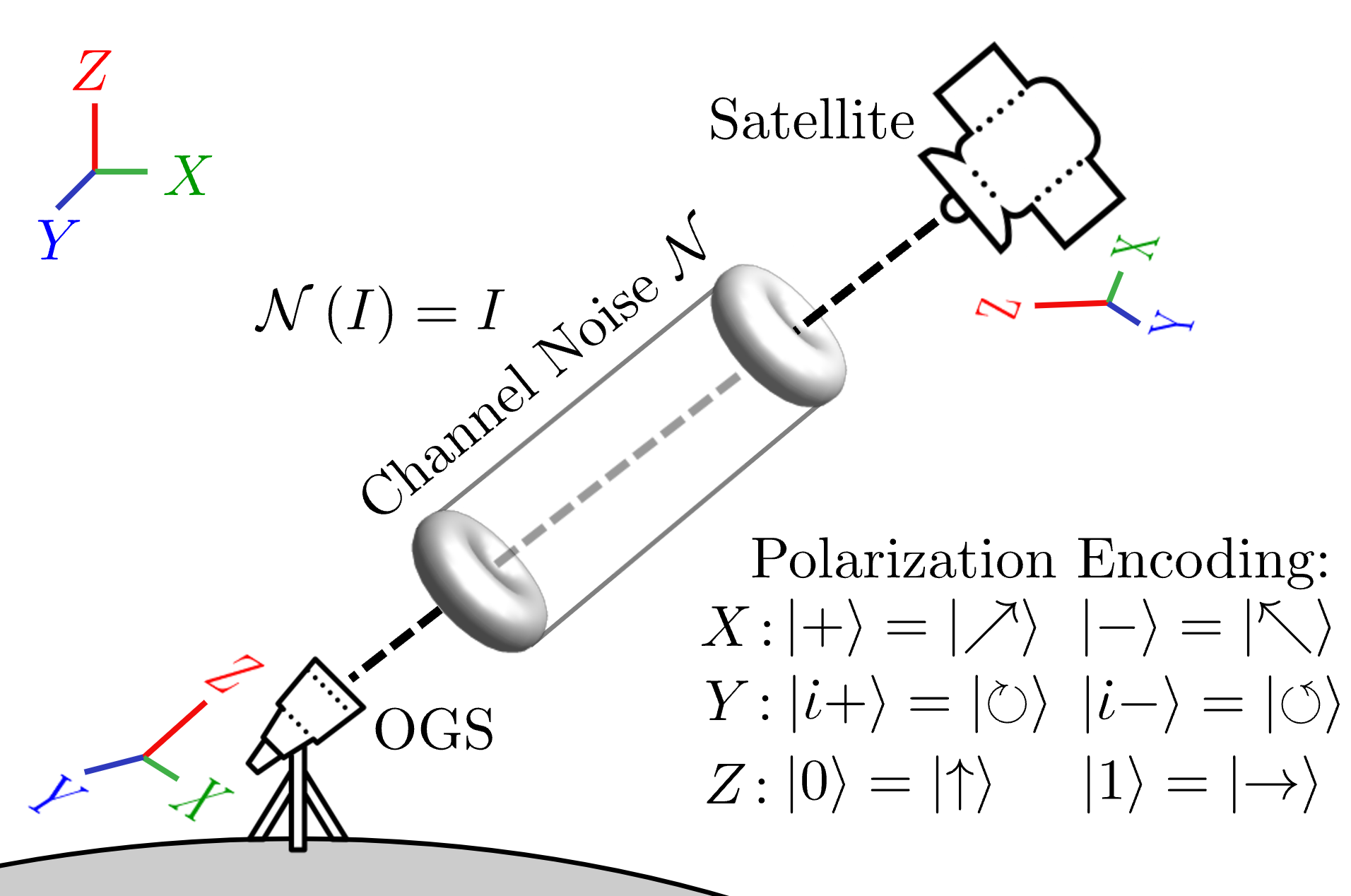}
	\caption{The system model. The transmitter and receiver lack a shared reference frame, which is exemplified as a satellite-to-ground quantum communication link where the local frames of reference do not coincide with the global reference. Polarization encoding of photons in global reference frame does not translate to the same in either of the local references. }
	\label{fig:systemmodel}
\end{figure}

The system model, exemplified by a satellite link, is illustrated in Fig.~\ref{fig:systemmodel}. 
We consider a transmitter, denoted by Alice, and a receiver, denoted by Bob, who are spatially separated and aim to perform \ac{QKD}. 
Due to practical limitations, such as relative motion, rotation, and imperfect polarization alignment, Alice and Bob do not share a common qubit reference frame. 
That is, the Pauli directions associated with their respective local systems do not coincide \cite{Bartlett:Clas:03, vanEnk:Quan:06}. 
The lack of a shared reference frame is a common practical challenge in polarization-encoded satellite quantum communication and path-encoded chip-to-chip quantum communication \cite{Laing:Refe:10}.

The communication link is assumed to be noisy and unknown. 
More specifically, over a given estimation and key-generation block, it is modeled as an unknown, stationary, unital qubit channel $\mc{N}$. 
The channel noise may possess a preferred basis that is not known to either party. 
For example, a Pauli-channel model typically specifies the error operators with respect to fixed Pauli axes, and therefore assumes that the Pauli error axes are known in the chosen reference frame \cite{Rehman:Enta:21}. In this work, this assumption is not made because Alice and Bob do not share a common qubit reference frame. The only structural assumption imposed on the qubit noise is unitality. The unital qubit channel represents the polarization/noise transformation conditioned on a successful detection event. Propagation loss, background counts, detector dark counts, and detector-efficiency mismatch are not modeled explicitly in the present asymptotic qubit-channel analysis. Consequently, the secret key rates reported here are expressed per sifted detected signal. Although the physical channel $\mc{N}$ and the local-frame unitaries $U_A$ and $U_B$ are unknown individually, Alice and Bob can estimate the statistics of the effective Alice-to-Bob channel from their preparation and measurement data. The optimized bases are then selected using this effective channel, without requiring a separate reconstruction of the physical noise process or the individual frame rotations.

\section{Optimizing the QKD Signaling}\label{sec:opt_QKD}

Mathematically, the lack of shared reference frame can be modeled by an unknown arbitrary but fixed rotation. More concretely, we assume that the global reference frame is fixed by the Pauli operators of  \eqref{eq:Pauli_operator} and the states shown at the bottom right in Fig.\ref{fig:systemmodel}. We  model the Alice's and Bob's rotated local reference frames by the unitaries $U_A$ and $U_B$, respectively. That is, when Alice prepares a state $\ket{\psi}_A$ locally, this state in the global reference is $U_A\ket{\psi}_A$. Similarly, when Bob locally measures a received state by the projector system $\left\{ \Pi_0^B, \Pi_1^B\right\}$, the measurement in the global reference is in fact $\left\{ U_B\Pi_0^B U_B^{\dagger}, U_B\Pi_1^BU_B^{\dagger}\right\}$. After the noisy transmission through $\QCN$, the probability of obtaining measurement outcome `0' is:
\begin{align}
	p_0 = \Tr{U_B\Pi_0^B U_B^{\dagger} \QCg{U_A\ketbra{\psi}_AU_A^{\dagger}}}.
\end{align}
In the case of $U_A$ and $U_B$ being known, Alice and Bob can compensate for the misalignment by applying $U_A^{\dagger}$ and $U_B^{\dagger}$ after local preparation and before locally measuring, respectively. In the case of $U_A = U_B = U$, but $U$ being unknown, they cannot recover the statistics of ideal system, i.e., all reference frames coincide with the global reference. Indeed, generally $\Tr{U\Pi_0 U^{\dagger} \QCg{U\ketbra{\psi}U^{\dagger}}} \neq \Tr{\Pi_0  \QCg{\ketbra{\psi}}}$ unless $\QCg{\cdot}$ is covariant with respect to $U$ \cite{Siudzinska:Quan:18}. In this work, $U_A \neq U_B$ and both being unknown is what we consider. In the following, we offer two approaches to perform \ac{QKD} under this setting.


\subsection{Approach 1: Direct Construction of \ac{PTM}}
The first approach amounts to first performing a direct construction of \ac{PTM}, followed by identifying an optimal set of mutually unbiased bases for \ac{QKD} that maximizes the sifted key rate. 

Since the Alice's and Bob's local reference frames do not match, it seems challenging to estimate the \ac{PTM}. We overcome this challenge by absorbing the unitaries $U_A$ and $U_B$ to extend the definition of channel from $\QCg{\cdot}$ to $\mc{U}^{\dagger}_B \circ \mc{N} \circ \mc{U}_A\left( \cdot\right)$. Essentially, we completely ignore the mismatched local reference frames and execute the \ac{PTM} construction protocol \cite{Roncallo:Paul:24}. 

In order to estimate the nontrivial ($i, j \neq 0$) entries of the total map between Alice and Bob, Alice prepares the eigenstate corresponding to the positive eigenvalue of \emph{her local} $P_j^{A}$ and sends it thought he channel. Due to the reference mismatch with the global reference the channel output is $\QCg{U_A \ketbra{\lambda_j}_A U^{\dagger}_A}$. Upon reception, Bob measures the Pauli operator $P_i^B = \Pi_{0,i}^B - \Pi_{1, i}^B$ in his local frame. Due to global mismatch, the measurement in global reference is $U_B P_i^B U_B^{\dagger} = U_B\Pi_{0,i}^BU_B^{\dagger} - U_B\Pi_{1, i}^BU_B^{\dagger}$. Thus, the entry $R_{i, j}$ that they measure is
\begin{align}
	R_{i, j}^{A \rightarrow B} &= \Tr{U_B P_i^B U_B^{\dagger} \QCg{U_A \ketbra{\lambda_j}_A U^{\dagger}_A}}\\
	&= \Tr{P_i^B\, U_B^{\dagger} \QCg{U_A \ketbra{\lambda_j}_B U^{\dagger}_A} U_B}\\
	&= \Tr{P_i^B\, \mc{U}^{\dagger}_B \circ \mc{N} \circ \mc{U}_A\left( \ketbra{\lambda_j}_A\right)}\\
	&= R_{i, j}^{\mc{U}^{\dagger}_B \circ \mc{N} \circ \mc{U}_A}.
\end{align}
The first equality is from Born's rule. The second equality is due to the cyclic property of trace. The third equality is defining the composition of $U_A$, $\QCN$, and $U_B^{\dagger}$ as a composite channel. At the end of \ac{PTM} estimation procedure, the communicating parties obtain the estimated \ac{PTM} $R^{\mc{U}^{\dagger}_B \circ \mc{N} \circ \mc{U}_A}$.

The matrix $R^{A \rightarrow B} = R^{\mc{U}^{\dagger}_B \circ \mc{N} \circ \mc{U}_A}$ defines the \ac{PTM} mapping from Alice's local reference frame followed by noise to the Bob's local reference frame. This matrix can be decomposed via the \ac{SVD} to obtain $R^{A \rightarrow B} = O_B \Sigma O_A$, where $O_B, O_A$ are the orthogonal matrices containing left and right singular vectors as their columns, and $\Sigma$ is the diagonal matrix containing the singular values of $R^{\mc{U}^{\dagger}_B \circ \mc{N} \circ \mc{U}_A}$. 

The \ac{SVD} of $R^{A \rightarrow B}$ simultaneously solves the challenge of absence of reference frame as well as that of the unknown preferred basis of the channel. This is a consequence of absorbing the lack of reference frame, cahracterized by $U_A$ and $U_B$ in the channel definition. Now, if Alice wants to transmit a state corresponding to the Bloch vector $\vec{r}$, she instead prepares the state corresponding to the Bloch vector $O_A^{\top}\vec{r}$. Similarly Bob, after receiving the channel output, transforms it by multiplying its Bloch vector with $O_B^{\top}$. Thus, the output Bloch vector of the channel is $\vec{s} = O_B^{\top} O_B \Sigma O_A O_A^{
\top} \vec{r} = \Sigma \vec{r}$. Thus, by employing the $O_A$ and $O_B$ at the transmitter and receiver, respectively, the effective channel noise and the lack of reference frame is reduced to the noise characterized by a diagonal \ac{PTM} $\Sigma$, i.e., a Pauli channel.

Due to the unitality of noise and the corresponding structure of $R^{A \rightarrow B}$, the first right and left singular vectors, corresponding to the singular value $\sigma_0 = 1$, are trivial and correspond to the Bloch vector of maximally mixed state. Furthermore, since the columns of orthogonal matrices are orthogonal (Bloch) vectors, they correspond to the elements of mutually unbiased basis of the two-dimensional Hilbert space. This observation gives us the following recipe of \ac{QKD} in the considered scenario of absence of shared reference frames and unital noise. $Z$ basis communication is defined by Alice preparing the second right singular vector in \emph{her local} reference and send it to Bob through the channel. Upon receiving, Bob measures the second left singular vector in \emph{his local} reference. This will result in error rate $Q_z = \frac{1  - \sigma_1}{2}$. The same procedure with the third and fourth singular vectors define communication in $X$ and $Y$ bases with corresponding error rates $Q_x = \frac{1  - \sigma_2}{2}$ and $Q_y = \frac{1  - \sigma_3}{2}$.\footnote{The labeling of $X$, $Y$, and $Z$ direction can be arbitrary. We choose $Z$, $X$, and $Y$ in the increasing order of \ac{QBER}. }

We follow the notation that $\sigma_1 \geq \sigma_2 \geq \sigma_3$. For the \ac{BB84} protocol, Alice and Bob utilize $X$ and $Z$ basis, which allows them the asymptotic key rate of \cite{Shor:Simp:00,Renner:SECU:08,  Murta:Key:20}
\begin{align}
	R_{\mathrm{BB84}} = 1 - h\left(Q_z\right) - h\left(Q_x\right),
\end{align}
bits per sifted detected signal, where we have defined the binary Shannon entropy $h\left(x\right) = -x \log_2 x - \left(1 - x\right) \log_2\left(1 - x\right)$. Similarly, for the six-state protocol where all three mutually unbiased bases are used, Alice and Bob may obtain the asymptotic key rate of \cite{Murta:Key:20}
\begin{align}
	R_{\text{six-state}} = 1 - H\left(\left\{ \lambda_i,j\right\}\right),
\end{align}
where $H\left( \left\{ p_i\right\}\right) = -\sum_i p_i \log_2 p_i$ is the Shannon entropy and 
\begin{align}
	\lambda_{0, 0} &= 1 - \frac{Q_x + Q_y + Q_z}{2}\\
	\lambda_{0, 1} &= \frac{Q_x + Q_y - Q_z}{2}\\
	\lambda_{1, 0} &= \frac{-Q_x + Q_y + Q_z}{2}\\
	\lambda_{1, 1} &= \frac{Q_x - Q_y + Q_z}{2}.
\end{align}

The optimality follows from the fact that (i) every unital channel is unitarily equivalent to a Pauli channel \cite{Puchala:Paul:19}, (ii)  the chosen bases choice has transformed the overall noise into the effective Pauli noise and (iii) the Pauli bases (where the \ac{PTM} is diagonal) are optimal for \ac{QKD} under Pauli noise \cite{Bae:Key:07}. 

\subsection{Approach 2: Sequential Basis Matching}

\begin{figure*}[!t]
	\centering
	\subfigure[~]{
	\includegraphics[width=0.31\textwidth]{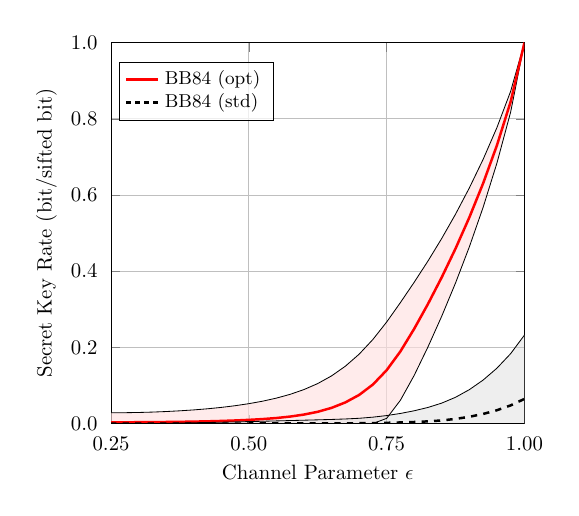}
	}
	\subfigure[~]{
		\includegraphics[width=0.31\textwidth]{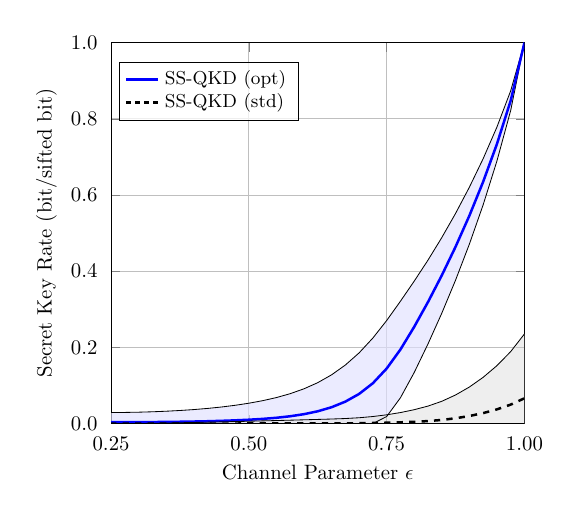}
	}
	\subfigure[~]{
		\includegraphics[width=0.31\textwidth]{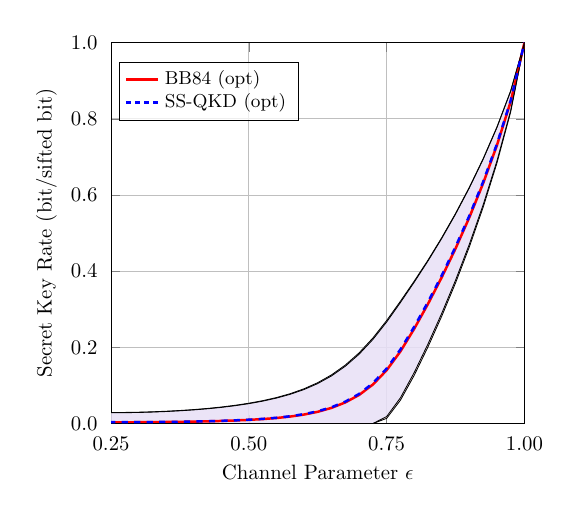}
	}
	\caption{Secret key rate (bits per sifted detected signal) as a function of channel parameter $\epsilon$. The optimized signaling, as proposed in the main text, compensates well for the absence of shared reference frame, as evidenced by the ideal secret key rate for the unitary/noiseless channel up to an unknown rotation, i.e., $\epsilon = 1$.}
	\label{fig:1optimizedqkdbb84}
\end{figure*}

The second approach that we propose here involves variational modeling of Alice's preparation and Bob's measurement directions and sequentially identifying the appropriate $Z$, $X$, and $Y$ directions for the \ac{QKD}. More concretely, an arbitrary pure qubit state can be parameterized by two angles $\theta \in \left[0, \pi\right]$ and $\phi \in \left[0, 2\pi \right]$ as
\begin{align}
	\ket{\psi} = \cos\frac{\theta}{2}\ket{0} + e^{i \phi}\sin \frac{\theta}{2}\ket{1}.
\end{align}
Alice and Bob utilize this parameterization and perturb these parameters variationally to identify the appropriate local directions that they locally define as $Z$. The proposed \ac{SBM} protocol operates as follows:
\begin{enumerate}
	\item \textbf{$Z$ Basis Search:} 
	\begin{enumerate}
		\item 
		Alice generates a parameter vector $\vec{\theta}^Z = \left[ \theta_A, \phi_A, \theta_B, \phi_B\right]$, prepares several copies of the state
		\begin{align}
			\ket{\psi}_A = \cos\frac{\theta_A}{2}\ket{0}_A + e^{i \phi_A}\sin \frac{\theta_A}{2}\ket{1}_A,
			\label{eq:Alice_variational_state_Z}
		\end{align}
		and sends them to Bob through the quantum channel. On the classical channel, she sends to Bob the parameter values $\theta_B$ and $\phi_B$.
		\item 
		Bob measures the received states with projectors $\left\{\Pi_0^B, I - \Pi_0^B\right\}$, where $\Pi_0^B = \ketbra{\psi}_B$ with 
		\begin{align}
			\ket{\psi}_B = \cos\frac{\theta_B}{2}\ket{0}_B + e^{i \phi_B}\sin \frac{\theta_B}{2}\ket{1}_B.
			\label{eq:Bob_variational_state_Z}
		\end{align}
		He obtains the outcome corresponding to $\Pi_0^B$ with probability $p_0 = \Tr{\Pi_0^B \QCg{\ketbra{\psi}_A}}$, which he can empirically estimate from the measurement outcomes. 
		\item Bob transmits the estimated $p_0$ to Alice, who employs an optimizer to maximize this values by varying the parameter vector appropriately.
		\item Alice and Bob repeat above three steps until they reach the maximum value of the estimated $p_0$. The identified directions in terms of their parameterized states define the $Z$ basis with the \ac{QBER}: $Q_z = 1 - p_0$. 
		\item
		They locally redefine $\ket{0}_A = \ket{\psi^*}_A$ and $\ket{0}_B = \ket{\psi^*}_B$, where $\ket{\psi^*}_A$ and $\ket{\psi^*}_B$ are the states \eqref{eq:Alice_variational_state_Z} and \eqref{eq:Bob_variational_state_Z}, respectively, with the optimal parameters. 
	\end{enumerate}
	
	\begin{figure*}[!t]
		\centering
		\subfigure[~]{
			\includegraphics[width=0.48\textwidth]{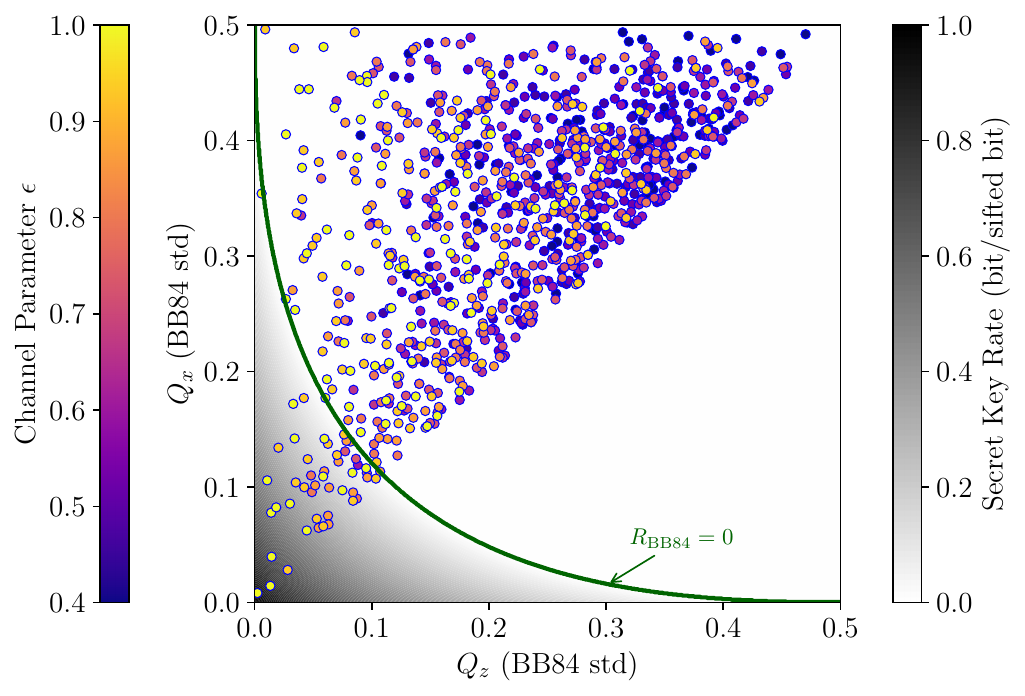}
		}
		\subfigure[~]{
			\includegraphics[width=0.48\textwidth]{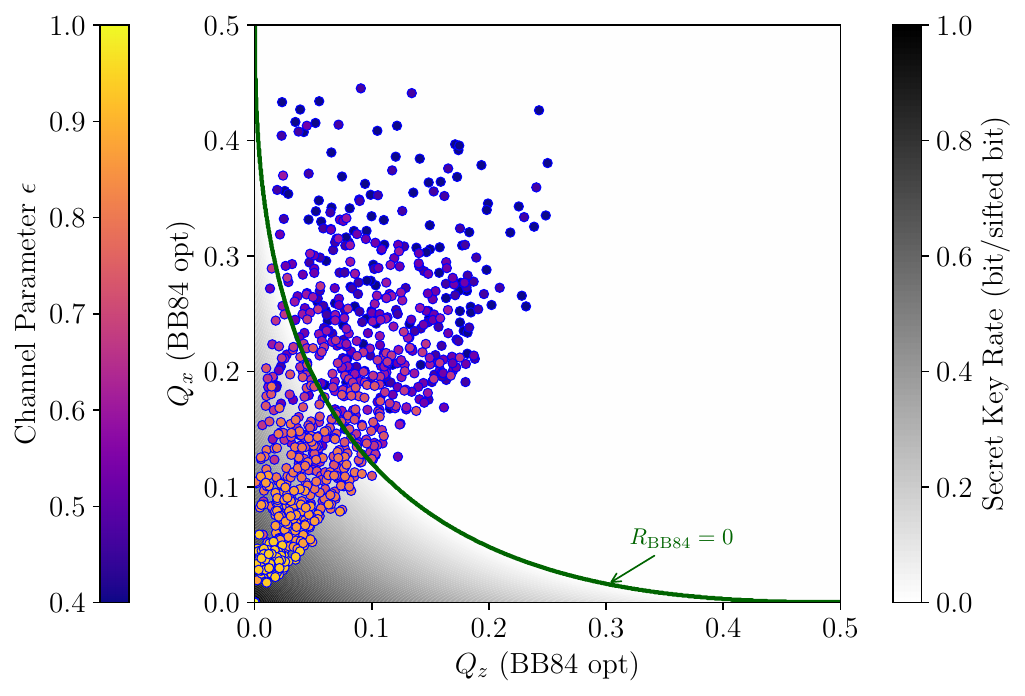}
		}
		\caption{Secret key rate (bits per sifted detected signal) as a function of QBERs in $Z$ and $X$ basis. The scatter points show the QBERs of numerical simulations with (a) standard, i.e., without basis optimization, and (b) optimized basis \ac{BB84} in random unitary channels. }
		\label{fig:bb84optcontour}
	\end{figure*}
	
	\item \textbf{$X$ Basis Search:}
	\begin{enumerate}
		\item 
		Alice generates another parameter vector $\vec{\theta}^X = \left[ \phi_A, \phi_B\right]$, prepares several copies of the state
		\begin{align}
			\ket{+}_A = \frac{\ket{0}_A + e^{i \phi_A}\ket{1}_A}{\sqrt{2}},
			\label{eq:Alice_variational_state_X}
		\end{align}
		and sends them to Bob through the quantum channel. On the classical channel, she sends to Bob the parameter value $\phi_B$.
		\item 
		Bob measures the received states with projectors $\left\{\Pi_+^B, I - \Pi_+^B\right\}$, where $\Pi_+^B = \ketbra{+}_B$ with 
		\begin{align}
			\ket{+}_B = \frac{\ket{0}_B + e^{i \phi_B}\ket{1}_B}{\sqrt{2}}.
			\label{eq:Bob_variational_state_X}
		\end{align}
		He obtains the outcome corresponding to $\Pi_+^B$ with probability $p_+ = \Tr{\Pi_+^B \QCg{\ketbra{+}_A}}$, which he can empirically estimate from the measurement outcomes. 
		\item Bob transmits the estimated $p_+$ to Alice, who employs an optimizer to maximize this values by varying the parameter vector appropriately.
		\item Alice and Bob repeat above three steps until they reach the maximum value of the estimated $p_+$. The identified directions in terms of their parameterized states define the $X$ basis with the \ac{QBER}: $Q_x = 1 - p_+$. 
		\item
		They locally redefine $\ket{+}_A = \ket{+^*}_A$ and $\ket{+}_B = \ket{+^*}_B$, where $\ket{+^*}_A$ and $\ket{+^*}_B$ are the states \eqref{eq:Alice_variational_state_X} and \eqref{eq:Bob_variational_state_X}, respectively, with the optimal parameters. 
	\end{enumerate}
	\item \textbf{$Y$ Basis Definition:}
	\begin{enumerate}
		\item Alice and Bob locally define
		\begin{align}
			\ket{i+}_A = \frac{\ket{0}_A + e^{i\left(\phi^*_A + \pi/2\right)} \ket{1}_A }{\sqrt{2}}
		\end{align}
		and
		\begin{align}
			\ket{i+}_B = \frac{\ket{0}_B + e^{i\left(\phi^*_B + \pi/2\right)} \ket{1}_B }{\sqrt{2}},
		\end{align}
		where $\phi^*_A$ and $\phi^*_B$ are the optimal parameters found in the $X$ basis search for defining the $X$ direction. 
		\item 
		Alice and Bob estimate the Y basis \ac{QBER} $Q_y$ by Alice sending several copies of $\ket{i+}_A$ and Bob measuring them with the projectors defined by the direction $\ket{i+}_B$.
	\end{enumerate}
\end{enumerate}

At the end of \ac{SBM}, Alice and Bob have local sets of mutually unbiased basis that they can use for \ac{BB84} or six-state \ac{QKD}. In particular, the achieved \ac{QBER} and the secret key rate by employing the \ac{SBM} is the same as the one obtained by the direct construction of the \ac{PTM}. This can be seen by recalling that it is not possible to obtain a lower \ac{QBER} than $Q_z = p_X + p_Y$ in a Pauli channel, thus the $Z$ basis search in \ac{SBM} by minimizing the \ac{QBER} results in the same $Q_z$ as the one obtained via the direct construction of the \ac{PTM}. The $X$ basis search is in the plane that is mutually unbiased to the identified $Z$ basis. This is equivalent to searching the orthogonal plane in the Bloch sphere. From Approach 1, we know that the lowest achievable \ac{QBER} in this plane is $Q_x$ of Approach 1. Thus, the optimization in \ac{SBM} should converge to the same value. Finally, fixing any two mutually unbiased bases in the two-dimensional Hilbert space automatically identifies the third basis. Thus, the achieved \ac{QBER} and the secret key rate by employing the \ac{SBM} is the same as the one obtained by the direct construction of the \ac{PTM}.

\section{Numerical Examples}\label{sec:NE}
In this section, we provide numerical examples of our work by simulating the system model and proposed approaches.\footnote{The simulation code is available under MIT License at the Github respository: \url{https://github.com/junaid572/QKD_No_Reference_Unital_Noise}.} We simulate the lack of shared reference frame by generating two Haar random unitaries $U_A$ and $U_B$ and rotate the Alice's and Bob's states/measurement with these, respectively. In order to simulate a unital channel, we simulate a random unitary channel
\begin{align}
	\QCg{\rho} = \sum_{i = 1}^4 p_i U_i \rho U_i^{\dagger},
\end{align}
where $U_i$ are Haar random unitaries and $p_1 = \epsilon$ and $ p_2 = p_3 = p_4 = \frac{1 - \epsilon}{3}$.

In Fig.~\ref{fig:1optimizedqkdbb84}, we plot the secret key rate (bits per sifted detected signal) for \ac{BB84} and six-state \ac{QKD} as a function of channel parameter $\epsilon$. In Fig.~\ref{fig:1optimizedqkdbb84}(a), we plot the secret key rate for \ac{BB84} with and without optimization. The solid red line shows the average secret key rate of $10^3$ runs for each $\epsilon$ value, and the red shaded region denotes one standard deviation of the data. The dashed black line shows the corresponding average for standard \ac{BB84}, i.e., without finding the optimal bases. We can see that the optimized signaling clearly outperforms the standard \ac{BB84} signaling, even for the unitary/noiseless channel up to an unknown rotation, i.e., $\epsilon = 1$. In Fig.~\ref{fig:1optimizedqkdbb84}(b), we plot the secret key rate for six-state \ac{QKD} on the same set of channels, where the solid blue line and blue shaded region denote the optimized average and its standard deviation, and the dashed black line denotes the standard six-state signaling. Fig.~\ref{fig:1optimizedqkdbb84}(c) compares the optimized secret key rates of the two protocols, with the solid red line for \ac{BB84} and the dashed blue line for six-state \ac{QKD}. We observe that the two \ac{QKD} protocols result in a very similar secret key rate. This is not the typical behavior in the symmetric \ac{QBER} case, where six-state \ac{QKD} always outperforms \ac{BB84}. This is a consequence of our modeling where we have assigned the $Y$ basis to be the most noisy. Thus, including the third basis in \ac{QKD} does not provide a significant added advantage.

\begin{figure*}[!t]
	\centering
	\subfigure[~]{%
		\includegraphics[height=0.268\textwidth]{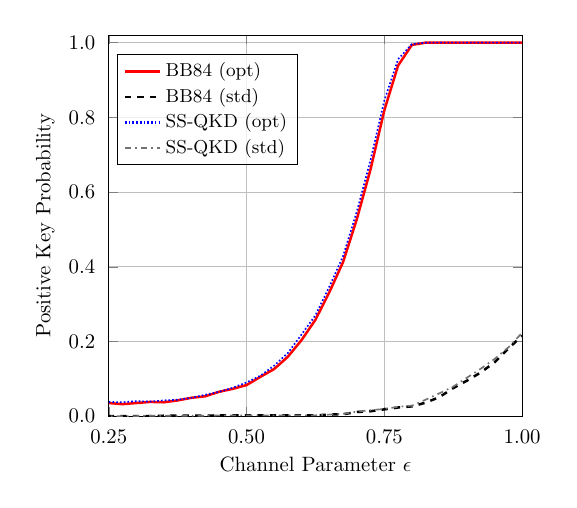}%
	}\hfill
	\subfigure[~]{%
		\includegraphics[height=0.268\textwidth]{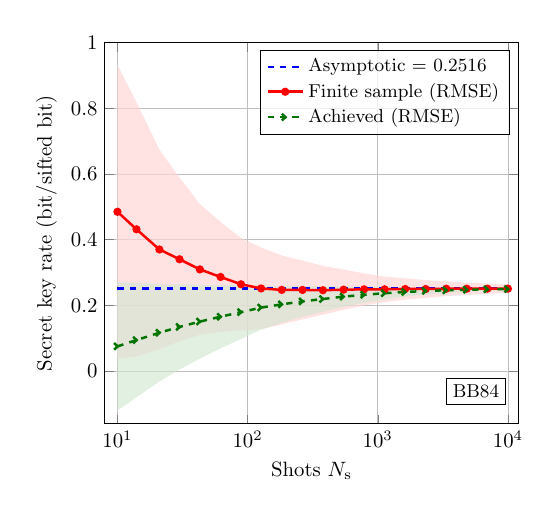}%
	}\hfill
	\subfigure[~]{%
		\includegraphics[height=0.268\textwidth]{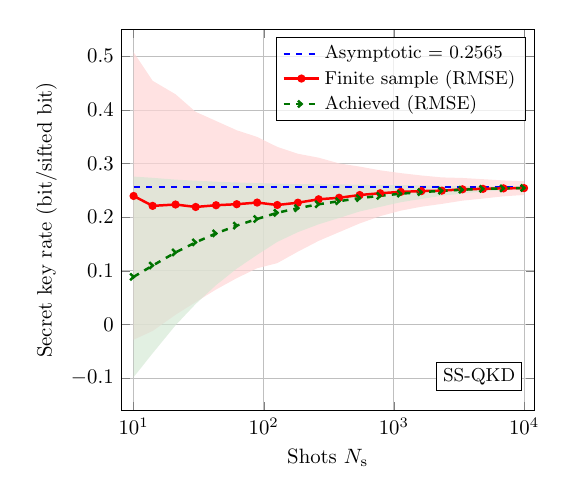}%
	}
	\caption{Robustness checks for the optimized signaling. 
		(a) Positive-key probability, $\Pr\{R>0\}$, over the random unitary channel ensemble as a function of the channel parameter $\epsilon$. 
		(b) Finite-sample \ac{PTM}-based basis estimation for \ac{BB84} at $\epsilon=0.8$, shown as a function of the number of shots per \ac{PTM} entry, $N_{\mathrm{s}}$. 
		(c) Corresponding finite-sample \ac{PTM}-based basis estimation for six-state \ac{QKD} at $\epsilon=0.8$. 
		In (b) and (c), the dashed line denotes the exact-\ac{PTM} asymptotic rate, the finite-sample curve denotes the rate computed from the noisy estimated \ac{PTM}, and the achieved-rate curve denotes the rate obtained when the finite-sample optimized bases are applied to the underlying simulated channel. 
		Shaded bands denote the root-mean-square error (RMSE).}
	\label{fig:robustnesschecks}
\end{figure*}

In the background (gray scale) of Fig.~\ref{fig:bb84optcontour}, we plot the asymptotic key exchange rate as a function of $Q_z$ and $Q_x$ for \ac{BB84} \ac{QKD}.  The green solid line shows the boundary outside of which (towards right) the secret key rate is zero. In the foreground of Fig.~\ref{fig:bb84optcontour}(a) and Fig.~\ref{fig:bb84optcontour}(b), we plot the achieved \ac{QBER} set $\left(Q_z, Q_x\right)$ for standard and optimized signaling \ac{QKD} with the same set of random unitary channels, respectively. For the standard \ac{BB84} in this setting, we observe that a majority of achieved \ac{QBER} lie outside the boundary of positive secret key rate. In other words, the majority of system realizations fail to achieve any \ac{QKD}. The proposed optimized signaling, however, manages to achieve a positive secret key rate for the same set of random unitary channels. These numerical examples demonstrate the effectiveness of the proposed approach for \ac{QKD} without shared reference frame and in the presence of unital noise. 

The robustness of the optimized signaling is further characterized in Fig.~\ref{fig:robustnesschecks}. 
Fig.~\ref{fig:robustnesschecks}(a) shows the positive-key probability, $\Pr\{R>0\}$, estimated over the random unitary channel ensemble as a function of the channel parameter $\epsilon$. This quantity measures the fraction of random channel realizations that support a strictly positive asymptotic secret key rate. For optimized signaling, the positive-key probability approaches unity as $\epsilon$ increases, whereas it remains low for standard local-basis signaling because the standard bases do not adapt to the effective reference-frame rotation or to the channel's preferred noise directions. Fig.~\ref{fig:robustnesschecks}(b) and Fig.~\ref{fig:robustnesschecks}(c) illustrate the effect of finite-sample \ac{PTM} estimation at $\epsilon=0.8$ for \ac{BB84} and six-state \ac{QKD}, respectively. Here, $N_{\mathrm{s}}$ denotes the number of measurement shots used to estimate each nontrivial \ac{PTM} entry. For each value of $N_{\mathrm{s}}$, the optimized bases are selected from the finite-sample estimate of the effective \ac{PTM}. The finite-sample key-rate estimate is computed from the \acp{QBER} inferred from this estimated \ac{PTM}, while the achieved rate is computed by applying the same finite-sample optimized bases to the underlying simulated channel.

For \ac{BB84}, the finite-sample key-rate estimate can be biased upward at small $N_{\mathrm{s}}$. This occurs because \ac{BB84} uses only two of the three optimized directions, so selecting the two most favorable estimated bases can make statistical fluctuations appear beneficial. This is a finite-sample selection effect, not an actual improvement of the underlying channel. The achieved-rate curve removes this optimistic estimation effect by evaluating the selected bases on the true simulated channel, and it converges to the exact-\ac{PTM} asymptotic rate as $N_{\mathrm{s}}$ increases.

For six-state \ac{QKD}, all three optimized bases enter the key-rate expression. 
Therefore, there is no analogous best-two-basis selection step, and finite-sample fluctuations affect all three basis-dependent \acp{QBER} jointly. In the simulated regime, the finite-sample key-rate estimate is therefore less biased upward and initially lies below the exact-\ac{PTM} asymptotic value. As $N_{\mathrm{s}}$ increases, both the finite-sample estimate and the achieved rate converge to the exact-\ac{PTM} optimized rate. These results demonstrate the statistical robustness of the proposed basis-estimation method, but they should not be interpreted as a finite-key security analysis.

\section{Conclusion}\label{sec:conclusion}
In this work, we investigated \ac{QKD} in the absence of a shared qubit reference frame, a condition that naturally arises in satellite and moving-platform quantum communication links. Rather than treating reference-frame alignment and channel-noise characterization as separate tasks, we formulated the problem through an effective Alice-to-Bob channel that jointly captures both effects. This perspective allows Alice and Bob to adapt their local signaling bases using only operationally accessible preparation and measurement statistics.

We proposed two approaches for this basis adaptation: a \ac{PTM}/\ac{SVD}-based method and an operational \ac{SBM} method. Both methods identify local mutually unbiased bases that align the observed correlations with the principal directions of the effective unital channel. In these optimized bases, the resulting asymmetric \acp{QBER} can be used directly in the asymptotic key-rate analysis of \ac{BB84} and six-state \ac{QKD}. Numerical results for random unitary channels show that such optimized signaling can substantially improve key-generation performance and can recover positive key rates in cases where standard local-basis signaling fails.

The results highlight the importance of treating basis choice as part of the QKD design problem when a shared reference frame is unavailable. The proposed framework provides a step toward more adaptive QKD implementations for dynamic quantum links, where both alignment and noise conditions may be difficult to characterize in advance. Future work may extend the approach beyond unital qubit channels and develop a finite-key security analysis that accounts for statistical uncertainty in the estimated channel parameters, optimized bases, and \acp{QBER}.

\balance
\bibliographystyle{IEEEtranDOI}
\bibliography{QKD_Channel_Estimation,LatexInclusion/IEEE-TIF-QRACIT}
\end{document}